\documentclass[cameraready]{Interspeech}
\usepackage[table]{xcolor}
\usepackage{cite}

\title{Audio-Cogito: Towards Deep Audio Reasoning in Large Audio Language Models}

\author[affiliation={1},]{Longhao}{Li}
\author[affiliation={2},]{Hongjie}{Chen}
\author[affiliation={2},]{Zehan}{Li}
\author[affiliation={2},]{Qihan}{Hu}
\author[affiliation={2},]{Jian}{Kang}
\author[affiliation={2}, correspondingauthor]{Jie}{Li}
\author[affiliation={1}, correspondingauthor]{Lei}{Xie}
\author[affiliation={2},]{Yongxiang}{Li}

\address{
$^{1}$ Audio, Speech and Language Processing Group (ASLP@NPU), School of Computer Science, \\
Northwestern Polytechnical University, Xi’an, China \\
$^{2}$ China Telecom Artificial Intelligence Technology (Beijing) Co., Ltd
}

\email{lhli@mail.nwpu.edu.cn, chenhj37@chinatelecom.cn, lxie@nwpu.edu.cn}


\usepackage{comment}
\usepackage{booktabs}
\usepackage{multirow}
\usepackage{arydshln}

\begin{document}

\maketitle

\begin{abstract}
    Recent advances in reasoning models have driven significant progress in text and multimodal domains, yet audio reasoning remains relatively limited. Only a few Large Audio Language Models (LALMs) incorporate explicit Chain-of-Thought (CoT) reasoning, and their capabilities are often inconsistent and insufficient for complex tasks. To bridge this gap, we introduce Audio-Cogito, a fully open-source solution for deep audio reasoning. We develop Cogito-pipe for high-quality audio reasoning data curation, producing 545k reasoning samples. Based on this dataset, we adopt a self-distillation strategy for model fine-tuning. Experiments on the MMAR benchmark, the only audio benchmark evaluating the CoT process, show that our model achieves the best performance among open-source models and matches or surpasses certain closed-source models in specific metrics. Our approach also ranks among the top-tier systems in the Interspeech 2026 Audio Reasoning Challenge.
    
\end{abstract}

\section{Introduction}
Recent advancements in Large Language Models (LLMs) have significantly boosted their capabilities, particularly through techniques like inference scaling and Chain-of-Thought (CoT). It has been widely demonstrated that CoT enhances reasoning effectively by decomposing complex queries into intermediate reasoning steps. This paradigm has successfully extended beyond text to multimodal systems, exemplified by visual reasoning models like LLaVA-Reasoner~\cite{LLaVa-Reasoner}.

In the audio processing community, audio-language modeling is also transitioning from foundational perception to complex cognitive reasoning. For example, recent Large Audio Language Models (LALMs)~\cite{salmonn, Qwen2-Audio, AudioFlamingo2, ltu, musilingo, mu-llama, gama, osum, mimo, step, kimi,gpt4o} and Omni Language Models (OLMs)~\cite{anygpt, openomni, baichuan, qwen25-omni, qwen3-omni, ming, omni-r1,gemini20flash} have made significant progress in speech perception and basic interaction. 
Meanwhile, Large Audio Reasoning Models (LARMs), including Audio-CoT~\cite{Audio-cot}, Audio Flamingo 3~\cite{AudioFlamingo3}, Step-Audio-R1~\cite{Step-Audio-R1}, and Qwen3-Omni-Thinking~\cite{qwen3-omni}, attempt to incorporate explicit CoT-style reasoning into the audio modality.

Despite these efforts, current LARMs still exhibit limited and unstable reasoning capabilities, as demonstrated by their performance on benchmarks like MMAR~\cite{mmar} and MMAU-Pro~\cite{mmau-pro}. A typical phenomenon is that these models often produce rigid and structured reasoning traces that lack deep audio grounding. Especially in complex acoustic environments, they remain susceptible to logical inconsistencies and the misinterpretation of subtle acoustic cues.
We attribute these limitations primarily to the scarcity of high-quality audio reasoning datasets. 
Current public audio datasets, such as AudioSet~\cite{audioset}, AudioCaps~\cite{audiocaps}, and Clotho~\cite{clotho}, typically provide brief labels or captions that are insufficient to cultivate deep audio reasoning.
While a handful of audio reasoning datasets exist~\cite{ Audio-reasoner, AudioFlamingo3}, they predominantly focus on shallow reasoning tasks. 
Furthermore, constructing datasets with complex reasoning traces relies heavily on closed-source models like Gemini 2.5 Pro~\cite{gemini2.5}. This reliance not only leads to substantial annotation costs and hinders reproducibility but also introduces incompatible inference formats across architectures, further constraining the practical applicability of existing resources.

To address these challenges, this study proposes \textbf{Audio-Cogito}\footnote{“Cogito” is Latin for “I think”}, a fully open-source solution that elicits deep audio reasoning capabilities in LALMs without reliance on proprietary APIs. 
We design \textbf{Cogito-Pipe}, a systematic pipeline for constructing high-quality audio reasoning datasets. The Cogito-Pipe consists of four stages, namely Data Collection, QA Construction, CoT Generation, and Quality Verification. During Data Collection, we aggregate diverse metadata across sound, speech, and music domains, followed by the synthesis of instruction pairs in the QA Construction stage.
Subsequently, we generate reasoning trajectories via self-distillation during the CoT generation stage. By employing the same model for both reasoning data generation and fine-tuning, we ensure consistency in reasoning patterns and mitigate performance degradation often caused by mismatched logic.
Finally, a dual-verification strategy ensures data quality and reliability. Experimental results on the MMAR benchmark demonstrate that Audio-Cogito achieves superior performance. Furthermore, Audio-Cogito secured top-tier performance in the Interspeech 2026 Audio Reasoning Challenge\cite{interspeech2026audioreasoning}, exhibiting strong capabilities in mixed-domains reasoning tasks.
\begin{figure*}[!t]
  \centering
  \includegraphics[width=\linewidth]{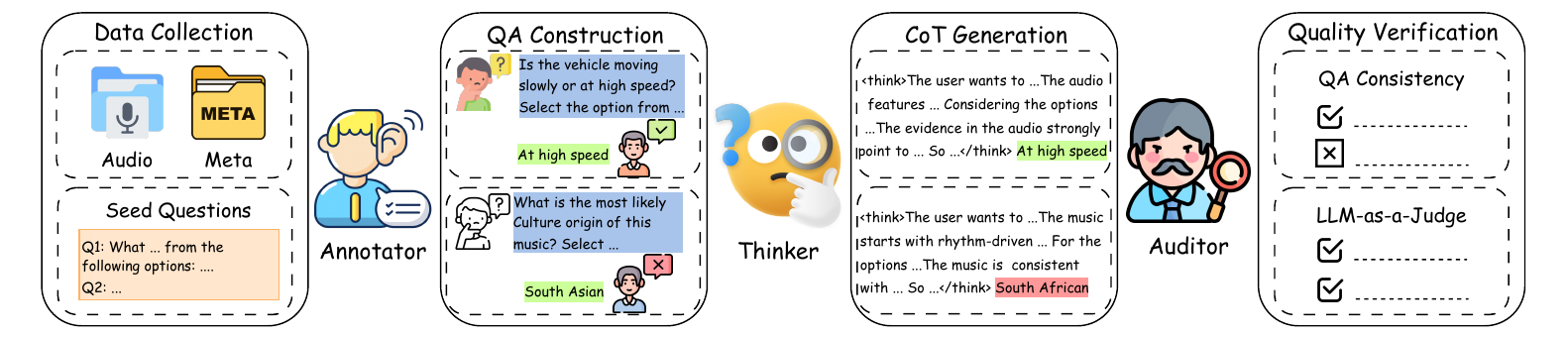}
  \caption{Overview of Cogito-Pipe.}
  \label{fig:data_pipline}
\end{figure*}
Our main contributions are: 
\begin{itemize}
\item We propose Audio-Cogito, built on Qwen3-Omni-Thinking, which utilizes self-distillation to substantially enhance the deep reasoning capabilities of LALMs.
\item We introduce Cogito-Pipe, a fully open-source four-stage pipeline for constructing high-quality and diverse audio reasoning data.
\item We release a large-scale audio reasoning dataset\footnote{https://github.com/llh666521/Audio-Cogito} with 545k high-quality samples spanning multiple audio domains.
\item Audio-Cogito achieves top-tier performance in the Interspeech 2026 Audio Reasoning Challenge and sets new SOTA results among open-source models on the MMAR benchmark, even surpassing several proprietary systems.
\end{itemize}

\section{Audio-Cogito}
\subsection{Cogito-Pipe}

In this section, we introduce our automated pipeline, Cogito-Pipe, to generate audio reasoning SFT data. As shown in Figure~\ref{fig:data_pipline}, the Cogito-Pipe consists of four stages: (1) Data Collection from multi-domain audio sources spanning sound, speech, and music; (2) QA Construction to synthesize diverse and challenging QA pairs; (3) CoT Construction to produce detailed step-by-step reasoning traces; and (4) Quality Verification to enforce consistency between QA pairs and CoT rationales while filtering out hallucinated or low-quality samples.

\subsubsection{Data Collection}
To construct a diverse, high-quality, and multi-task audio reasoning dataset, we extensively collect audio samples across three primary audio domains: sound events, speech, and music, including scenarios with mixed or interleaved audio domains. We collect the associated metadata to provide supplementary descriptive context for the audio samples. Furthermore, we curate a seed question pool of approximately 500 high-quality questions as few-shot exemplars to guide the generation of diverse, challenging, and reasoning-oriented data during QA Construction. This seed question pool is built through a collaborative pipeline of LLM generation and expert refinement. Specifically, we first use an LLM to generate candidate questions spanning multiple audio domains, reasoning types, and difficulty levels. These candidates are then reviewed, revised, and supplemented by domain experts to produce the final set of curated seed questions.

\begin{table}[b]
\centering
\scriptsize
\setlength{\tabcolsep}{2pt}
\caption{Statistics of the datasets used in Cogito-Pipe.}
\label{tab:dataset_stats}
\begin{tabular}{@{}lllccc@{}}
\toprule
\textbf{Domain} & \textbf{Dataset Source} & \textbf{Main Skills Learning} & \textbf{Quantity} & \textbf{Ratio (\%)} \\ \midrule
\multirow{4}{*}{Sound}  & AudioSet~\cite{audioset}       & General Audio Event  & 179k & 32.53 \\
                        & Clotho~\cite{clotho}         & Audio Captioning     & 6k   & 1.14  \\
                        & AudioCaps~\cite{audiocaps}      & Audio Captioning     & 40k  & 7.20  \\
                        & ComplexAudio~\cite{Audio-reasoner}   & Complex Audio        & 37k  & 6.66  \\ \midrule
\multirow{3}{*}{Speech} & MELD~\cite{meld}           & Speech Emotion       & 24k  & 4.50  \\
                        & CoVoST2~\cite{s2tt}        & Speech Translation   & 56k  & 10.10 \\
                        & DailyTalk~\cite{dailytalk}      & Spoken Dialogue      & 9k   & 1.64  \\ \midrule
\multirow{3}{*}{Music}  & MusicBench~\cite{musicbench}     & General Music        & 88k  & 16.04 \\
                        & FMA~\cite{fma}           & Music Genre          & 76k  & 13.81 \\
                        & Medley-solos-DB~\cite{medleysolosdb} & Instrument Analysis  & 35k  & 6.38  \\ \bottomrule
\end{tabular}
\end{table}

\subsubsection{QA Construction}
We employ Qwen3-Omni-Instruct as the annotator for QA construction. To enhance both quality and diversity, for each QA construction instance, we sample 20 questions from the pre-constructed seed question pool and use them as few-shot exemplars.. This guides the model to mimic specific questioning styles and perspectives, thereby facilitating the extraction of in-depth auditory knowledge. 
Furthermore, we explicitly instruct the model to generate confusing distractor options as hard negatives. For each audio clip, 1-3 QA pairs are generated, ensuring that a wide variety of questions can capture auditory cues from multiple angles.

\subsubsection{CoT Generation}
We employ Qwen3-Omni-Thinking as the thinker to generate reasoning chains via a self-distillation strategy, where the identical model architecture is utilized for both data generation and the subsequent fine-tuning phase. 
Specifically, we adopt a free-form CoT generation strategy, allowing model outputs to deviate from rigid templates. 
Our empirical experiments suggest that the format misalignment between rigid templates and the model's native output patterns degrades its intrinsic reasoning capabilities. Furthermore, although the ground-truth answers are available, we deliberately withhold them during generation. This forces the model to derive answers solely from acoustic cues, ensuring that its reasoning process remains faithful to the audio input.

\subsubsection{Quality Verification}
To guarantee the high quality of the generated audio reasoning data, we implement an auditor, a two-stage quality verification mechanism. First, we perform a QA Consistency Check to validate whether the answer derived from the CoT aligns with the answer in the constructed QA pairs. Subsequently, we employ an LLM-as-a-Judge paradigm using Qwen3-Omni-Instruct to scrutinize the reasoning process, explicitly filtering out samples that exhibit hallucinations or logical inconsistencies. 

Consequently, through the four stages of Cogito-Pipe, we obtain diverse and high-quality audio reasoning data. Furthermore, the models within the pipeline are interchangeable, allowing for self-distillation data generation by utilizing the specific target model intended for training.

\subsection{Model Training}
In Audio-Cogito, each input consists of an audio signal $A$ and a textual query $Q$, which are integrated into a multimodal input representation.
We explicitly decompose the model's generation into two parts: a Chain-of-Thought (CoT) reasoning trace $C$ that records step-by-step deductions, and a final response $R$ that provides the concluding answer. Accordingly, the model is trained to generate the concatenated sequence $(C, R)$, which we model with:
\begin{equation}
  P(C, R \mid A, Q; \theta) = f_{\theta}(A, Q).
\end{equation}
To enable explicit learning of both reasoning and answer generation, we construct a dataset:
\begin{equation}
  \mathcal{D} = \{(A_i, Q_i, C_i, R_i)\}_{i=1}^{N}
\end{equation}
where each sample contains the audio input $A_i$, the corresponding query $Q_i$, the structured reasoning trace $C_i$, and the final answer $R_i$. This formulation encourages the model to learn structured, logically grounded responses.

Training maximizes the joint likelihood of $C$ and $R$, encouraging the model to reason before producing the final answer. The objective is defined as:
\begin{equation}
  \mathcal{L}(\theta) = -\sum_{i=1}^{N} \log P(C_i, R_i \mid A_i, Q_i; \theta).
\end{equation}
Optimizing this objective trains Audio-Cogito to articulate an explicit reasoning process prior to delivering the final outcome, improving interpretability and reliability while better aligning model behavior with human-style problem solving.

\section{Experiments}
\subsection{Experimental Setup}

\subsubsection{Training Details}
Our model, \texttt{Audio-Cogito}, is built upon the \texttt{Qwen3-Omni-Thinking} with 30 billion parameters. We utilize the ms-swift\footnote{https://github.com/modelscope/ms-swift} framework to conduct supervised fine-tuning using Low-Rank Adaptation (LoRA). The model is fine-tuned for one epoch on the dataset constructed via Cogito-Pipe, with a maximum learning rate set to $1 \times 10^{-5}$.

\subsubsection{Evaluation Metrics} 
 Conventional audio benchmarks~\cite{audiobench, airbench, mmau, mmsu, televal, urobench} predominantly rely on final-answer accuracy as the sole performance metric. Such outcome-oriented evaluation often masks whether a model arrives at the correct answer through sound logic or spurious correlations. By contrast, \textsc{MMAR}~\cite{mmar} establishes a standardized protocol explicitly dedicated to evaluating the intermediate reasoning process, fostering a new direction for explainable audio intelligence. Thus, we exclusively employ the \textsc{MMAR} dataset as our evaluation benchmark. 

We adopt the evaluation protocol of the Interspeech 2026 Audio Reasoning Challenge\footnote{https://audio-reasoning-challenge.github.io/} to assess both answer correctness and reasoning quality. 
Specifically, for each sample $i$, let $c_i \in \{0,1\}$ denote the correctness of the answer, where $c_i=1$ indicates a correct prediction and $c_i=0$ otherwise. The answer's correctness is measured by the average accuracy (Avg) over the dataset: 
\begin{equation}
    \text{Avg} = \frac{1}{N}\sum_{i=1}^{N} c_i
\end{equation}
where $N$ is the total number of evaluation samples. 

Each MMAR sample is associated with an instance-level rubric, automatically generated by \texttt{Gemini-2.5-Pro} from the ground-truth reasoning path. The rubric contains five verifiable criteria that capture the key reasoning steps for that specific example. Given a model's predicted reasoning trace, an LLM judge evaluates whether each criterion is satisfied. Following the official challenge protocol, we use \texttt{GPT-4o} as the LLM judge. For a correctly answered sample, the judge assigns a binary score (0 or 1) to each criterion, and the reasoning score $r_i$ is computed as the proportion of satisfied criteria:
\begin{equation}
r_i = \frac{\text{\# satisfied rubric items}}{\text{\# total rubric items}}
\end{equation}
If the final answer is incorrect ($c_i=0$), the reasoning score is set to $r_i=0$. The overall Rubrics Score across the dataset is defined as:
\begin{equation}
\text{Rubrics} = \frac{1}{N}\sum_{i=1}^{N} r_i,
\end{equation} 
where $r_i$ takes values in $\{0, 0.2, 0.4, 0.6, 0.8, 1.0\}$ for correct predictions, and 0 otherwise.
We further introduce \textbf{Correct Reasoning Score (CRS)} to evaluate reasoning quality on the correct answer only as follows:
    \begin{equation}
    \text{CRS} = \frac{\sum_{i=1}^{N} r_i}{\sum_{i=1}^{N} c_i}
    \end{equation}
CRS can be interpreted as the average reasoning score conditioned on correct answers, providing a complementary view of reasoning quality.
To reduce evaluation variance, we conduct five runs and report the mean of the middle three scores.

\subsubsection{Baseline Models} 
We evaluate three categories of audio-capable models, using representative models from each category for comparison. (1) Large audio language models (LALMs), primarily designed for audio--text understanding, including open-source models such as \texttt{Audio Flamingo 2}~\cite{AudioFlamingo2} and \texttt{Qwen2-Audio-Instruct}~\cite{Qwen2-Audio}, as well as proprietary systems including \texttt{Omni-R1}~\cite{omni-r1} and \texttt{GPT-4o Audio}~\cite{gpt4o}. (2) Omni language models (OLMs), which support fully multimodal input and output, covering open-source models such as \texttt{Qwen2.5-Omni}~\cite{qwen25-omni} and \texttt{Qwen3-Omni-Instruct}~\cite{qwen3-omni}, alongside proprietary models including \texttt{Gemini 2.0 Flash}~\cite{gemini20flash} and \texttt{Gemini 2.5 Pro}~\cite{gemini2.5}. (3) Large audio reasoning models (LARMs), which extend LALMs by incorporating explicit Chain-of-Thought reasoning mechanisms, including models such as \texttt{Step-Audio-R1}~\cite{Step-Audio-R1} and \texttt{Qwen3-Omni-Thinking}~\cite{qwen3-omni}.

\subsection{Main Results}

\begin{table*}[htbp]
\centering
\scriptsize
\setlength{\tabcolsep}{1.5pt}
\renewcommand{\arraystretch}{0.95}

\caption{MMAR results across three model categories: LALMs, OLMs, and LARMs. The best-performing models within each category are highlighted in bold, and the second-best results are underlined. Dashed lines separate open-source and proprietary models.}
\label{tab:main_results}
\resizebox{\textwidth}{!}{
\begin{tabular}{l c ccc cccc c c c}
\toprule
\multirow{2}{*}{Models} & \multirow{2}{*}{Size}
& \multicolumn{3}{c}{Single Domain (\%)}
& \multicolumn{4}{c}{Mixed Domains (\%)}
& \multirow{2}{*}{Avg (\%)}
& \multirow{2}{*}{Rubrics (\%)}
& \multirow{2}{*}{CRS} \\
\cmidrule(lr){3-5} \cmidrule(lr){6-9}
& & Sound & Music & Speech
& Sound-Music & Sound-Speech & Music-Speech & Sound-Music-Speech
& & & \\
\midrule
Random Guess & -
& 29.39 & 25.88 & 31.48
& 25.00 & 29.30 & 31.10 & 28.13
& 29.32 & - & - \\
\midrule
\multicolumn{11}{c}{\textbf{Large Audio Language Models (LALMs)}} \\
\midrule
SALMONN~\cite{salmonn} & 7B & 30.90 & 29.60 & 34.40 & 9.10 & 37.60 & 28.10 & 37.50 & 32.80 & - & - \\
Audio Flamingo~\cite{AudioFlamingo} & 2.2B & 32.70 & 21.80 & 24.80 & 18.20 & 30.30 & 24.40 & 25.00 & 26.60 & - & - \\
Audio Flamingo 2~\cite{AudioFlamingo2} & 3B & 24.90 & 17.50 & 20.80 & 18.20 & 26.60 & 23.20 & 8.30 & 21.90 & - & - \\
Qwen2-Audio~\cite{Qwen2-Audio} & 8.4B & 33.90 & 23.30 & 33.00 & 9.10 & 33.00 & 26.80 & 33.30 & 30.40 & - & - \\
Qwen2-Audio-Instruct~\cite{Qwen2-Audio} & 8.4B & 33.30 & 24.30 & 32.30 & 9.10 & 31.20 & 30.50 & 25.00 & 30.00 & - & - \\
\hdashline
GPT-4o mini Audio~\cite{gpt4o} & - & 38.80 & 35.90 & 58.80 & \underline{45.50} & 60.10 & 57.30 & 60.00 & 50.60 & - & - \\
Omni-R1~\cite{omni-r1} & 8.4B & \textbf{67.30} & \textbf{51.50} & \underline{64.30} & \underline{45.50} & \underline{70.20} & \textbf{64.60} & \underline{70.80} & \underline{63.40} & - & - \\
GPT-4o Audio~\cite{gpt4o} & - & \underline{53.90} & \underline{51.00} & \textbf{70.40} & \textbf{63.60} & \textbf{72.50} & \underline{62.20} & \textbf{75.00} & \textbf{63.50} & - & - \\
\midrule
\multicolumn{11}{c}{\textbf{Omni Language Models (OLMs)}} \\
\midrule
Qwen2.5-Omni~\cite{qwen25-omni} & 10.7B & 58.80 & 40.80 & 59.90 & 54.50 & 61.90 & 67.10 & 58.30 & 56.70 & - & - \\
Qwen3-Omni-Instruct~\cite{qwen3-omni} & 30B-A3B & 59.39 & \underline{54.37} & 72.45 & 63.67 & \underline{77.52} & 65.85 & 66.67 & 66.90 & - & - \\
\hdashline
Gemini 2.0 Flash~\cite{gemini20flash} & - & \underline{61.20} & 51.00 & 72.10 & \underline{81.80} & 72.50 & 65.90 & \underline{70.80} & 65.60 & - & - \\
Gemini 2.5 Flash~\cite{gemini2.5} & - & 60.00 & 53.40 & \underline{77.20} & 63.60 & 76.20 & \underline{69.50} & \textbf{75.00} & \underline{68.40} & - & - \\
Gemini 2.5 Pro~\cite{gemini2.5} & - & \textbf{67.30} & \textbf{56.80} & \textbf{82.00} & \textbf{100.00} & \textbf{84.90} & \textbf{80.50} & 66.70 & \textbf{74.40} & - & - \\
\midrule
\multicolumn{11}{c}{\textbf{Large Audio Reasoning Models (LARMs)}} \\
\midrule
Mellow~\cite{mellow} & 167M & 33.30 & 26.70 & 24.80 & 18.20 & 37.20 & 32.90 & 29.20 & 30.00 & 18.50 & 0.61 \\
Audio-CoT~\cite{Audio-cot} & 8.4B & 35.80 & 25.20 & 34.00 & 9.10 & 30.70 & 30.50 & 37.50 & 31.30 & 19.85 & 0.62 \\
Audio-Reasoner~\cite{Audio-reasoner} & 8.4B & 42.42 & 32.52 & 42.52 & 45.13 & 48.62 & 30.49 & 29.17 & 40.50 & 28.40 & 0.68 \\
Audio Flamingo 3~\cite{AudioFlamingo3} & 7B & 56.97 & 45.15 & 59.52 & 45.45 & 67.89 & 59.76 & 41.67 & 57.40 & 45.20 & 0.79 \\
Step-Audio-R1~\cite{Step-Audio-R1} & 33B & 32.52 & 32.52 & \underline{68.71} & 45.22 & 72.02 & 62.20 & \underline{72.02} & 58.60 & 46.55 & 0.79 \\
Qwen3-Omni-Thinking~\cite{qwen3-omni} & 30B-A3B & \underline{64.24} & \underline{50.00} & \textbf{79.25} & \underline{54.55} & \underline{72.48} & \underline{69.51} & 70.83 & \underline{68.00} & \underline{57.97} & \underline{0.85} \\
\rowcolor{green!15} Audio-Cogito & 30B-A3B & \textbf{66.67} & \textbf{53.40} & \textbf{79.25} & \textbf{90.91} & \textbf{79.90} & \textbf{76.83} & \textbf{79.17} & \textbf{71.70} & \textbf{62.22} & \textbf{0.87} \\
\bottomrule
\end{tabular}
}
\end{table*}

Table~\ref{tab:main_results} presents the performance of three model categories, including the proposed \texttt{Audio-Cogito}, on MMAR under both single-domain and mixed-domains settings. 
The evaluation covers a total of seven subcategories across these conditions, with average accuracy reported over all subcategories. Performance is further assessed using the Rubrics Score (Rubrics) and the Correct Reasoning Score (CRS). Both open-source and closed-source models are included in the comparison.

As shown in Table \ref{tab:main_results}, \texttt{Audio-Cogito} achieves SOTA performance among open-source LARMs, LALMs, and OLMs on the MMAR benchmark. It attains the best average accuracy among the compared open-source models, surpassing \texttt{Qwen3-Omni-Thinking} by $5.44\%$ in relative terms. The gains are especially notable on mixed-domain tasks, further demonstrating the superior reasoning ability of \texttt{Audio-Cogito} in complex acoustic environments.

\texttt{Audio-Cogito} also narrows the gap between open-source and proprietary systems. Specifically, \texttt{Audio-Cogito} surpasses the average accuracy of closed-source OLMs such as \texttt{Gemini 2.0 Flash} and \texttt{Gemini 2.5 Flash}, alongside leading LALMs like \texttt{Omni-R1} and \texttt{GPT-4o Audio}. Compared with the current SOTA model, \texttt{Gemini 2.5 Pro}, \texttt{Audio-Cogito} even achieves better performance in Sound-Music-Speech and comparable performance in Single Domain Sound. These results show that \texttt{Audio-Cogito} approaches the performance of top-tier proprietary models in audio reasoning.

Beyond raw accuracy, \texttt{Audio-Cogito} demonstrates superior reasoning quality on the MMAR benchmark, as evidenced by the reasoning quality metrics Rubrics and CRS. As shown in Table \ref{tab:main_results}, our model achieves the best Rubrics and CRS among all LARMs, surpassing strong baselines such as \texttt{Qwen3-Omni-Thinking} and \texttt{Step-Audio-R1}. This indicates that \texttt{Audio-Cogito} produces highly reliable reasoning chains when generating correct answers, reflecting its stronger reasoning quality. These results further validate the effectiveness of our self-distillation strategy in fostering deep and logically grounded reasoning.

\subsection{Ablation Study}
To investigate the contribution of each stage in Cogito-Pipe, we fine-tune \texttt{Qwen3-Omni-Thinking} on datasets with specific components removed. As shown in Table \ref{tab:ablation_study}, all ablation configurations lead to performance degradation, validating the effectiveness of the proposed data construction pipeline. Specifically, removing seed questions results in the largest performance drop, particularly in mixed-domains tasks, indicating that seed questions introduce challenging and diverse queries that stimulate deeper reasoning. Removing quality verification significantly increases hallucinations, highlighting its role in maintaining dataset quality. Excluding meta information reduces QA accuracy by removing key grounding cues necessary for precise supervision. Overall, these components work together to enable Cogito-Pipe to construct high-quality reasoning data, allowing \texttt{Audio-Cogito} to surpass the base model.

\begin{table}[htbp]
\centering
\caption{Ablation study of Audio-Cogito on MMAR. S-M, S-S and M-S denote Sound-Music, Sound-Speech, and Music-Speech, respectively; S-M-S denotes Sound-Music-Speech.}
\label{tab:ablation_study}
\renewcommand{\arraystretch}{1.1} 
\setlength{\tabcolsep}{2pt}      
\resizebox{\columnwidth}{!}{
    \begin{tabular}{lcccccccccc}
    \toprule
    \textbf{Models} & \textbf{Sound} & \textbf{Music} & \textbf{Speech} & \textbf{S-M} & \textbf{S-S} & \textbf{M-S} & \textbf{S-M-S} & \textbf{Avg (\%)} & \textbf{Rubrics} & \textbf{CRS} \\
    \midrule
    Audio-Cogito & 66.06 & \textbf{53.40} & 79.25 & \textbf{90.91} & \textbf{77.98} & \textbf{75.61} & \textbf{75.06} & \textbf{71.20} & \textbf{62.22} & \textbf{0.87} \\
    \quad w/o seed questions & \textbf{66.67} & 50.97 & \textbf{79.59} & 72.73 & 73.39 & 69.51 & 62.50 & 68.90 & 58.80 & 0.85 \\
    \quad w/o quality verification & 64.85 & 52.43 & 78.57 & 81.82 & 76.61 & 73.17 & 70.83 & 69.90 & 60.40 & 0.86 \\
    \quad w/o meta information & 65.45 & 52.91 & 78.91 & \textbf{90.91} & 77.52 & 74.39 & 70.83 & 70.60 & 61.80 & 0.87 \\

    \bottomrule
    \end{tabular}
}
\end{table}

\section{Conclusion}

In this work, we introduce Audio-Cogito, an open-source solution for deep audio reasoning in LALMs. Leveraging Cogito-Pipe for high-quality data curation, we construct and release a 545k-sample open-source audio reasoning dataset. We further employ a self-distillation strategy that enhances complex reasoning capabilities. Experiments on the MMAR benchmark show that Audio-Cogito achieves SOTA performance among open-source models and narrows the gap with leading proprietary systems, while its top-tier performance in the Interspeech 2026 Audio Reasoning Challenge further validates its effectiveness. High Rubrics and CRS scores also indicate that our approach produces reliable Chain-of-Thought processes. 

\newpage
\section{Generative AI Use Disclosure}
Generative AI tools were employed exclusively for linguistic refinement and editorial assistance. These tools were not used to develop the methodology, conduct the experiments, generate the results, or draw the conclusions of this work. The authors retain full responsibility and accountability for all aspects of the manuscript.

\bibliographystyle{IEEEtran}
\bibliography{mybib}

\end{document}